\documentclass[prb,amsmath,amssymb,showkeys,floatfix,prb,longbibliography]{revtex4-1}
\usepackage{graphicx,psfrag}
\usepackage{dcolumn}
\usepackage{bm,bbm}

\begin{document}

\title{Scattering from a quantum anapole at low energies}

\author{Kyle M.~Whitcomb}

\author{David C.~Latimer}

\affiliation{Department of Physics, University of Puget Sound,
Tacoma, WA 98416-1031
}

\newcommand*{\sech}{\mathop{\mathrm{sech}}\limits}
\newcommand*{\balpha}{\boldsymbol{\alpha}} 
\newcommand*{\dilog}{\mathrm{Li}_2}
\newcommand{\qslash}{\not{\hbox{\kern-2pt $q$}}}
\newcommand{\kslash}{\not{\hbox{\kern-2pt $k$}}}
\newcommand{\pslash}{{\not{\hbox{\kern -2pt $p$}}}}
\newcommand{\delslash}{\not{\hbox{\kern-3pt $\partial$}}}
\newcommand{\Dslash}{\not{\hbox{\kern-3pt $D$}}}
\newcommand{\gmn}{g^{\mu \nu}}
\newcommand{\Pslash}{\not{\hbox{\kern-2.3pt p}}}
\newcommand{\Kslash}{\not{\hbox{\kern-2.3pt $K$}}}
\newcommand{\Pslashsup}{^\not{\hbox{\kern-0.5pt $^P$}}}
\newcommand{\Poddup}{^\not{\hbox{\kern-0.5pt $^\mathcal{P}$}}}
\newcommand{\Podddown}{_\not{\hbox{\kern-0.5pt $_\mathcal{P}$}}}
\newcommand{\bsig}{\boldsymbol{\sigma}}
\newcommand{\beps}{\boldsymbol{\epsilon}}
\newcommand{\phat}{\hat{\mathbf{p}}}
\newcommand{\khat}{\hat{\mathbf{k}}}
\newcommand{\al}[1]{\begin{align}#1\end{align}}
\newcommand{\diff}{\mathrm{d}}

\begin{abstract}

In quantum field theory, the photon-fermion vertex can be described in terms of four form factors which encode the static electromagnetic properties of the particle, namely its charge, magnetic dipole moment, electric dipole moment, and anapole moment.     For Majorana fermions, only the anapole moment can be nonzero, a consequence of the fact that these particles are their own antiparticles.  Using the framework of quantum field theory, we perform a scattering calculation which probes the anapole moment with a spinless charged particle.  In the limit of low-momentum transfer, we confirm that the anapole can be classically likened to a point-like toroidal solenoid whose magnetic field is confined to the origin.  Such a toroidal current distribution can be used to demonstrate the Aharonov-Bohm effect.  We find that, in the non-relativistic limit, our scattering cross section agrees with a quantum mechanical computation of the cross section for a spinless current scattered by an infinitesimally thin toroidal solenoid.  Our presentation is geared toward advanced undergraduate or beginning graduate students. This work serves as an introduction to the anapole moment and also provides an example of how one can develop an understanding of a particle's electromagnetic properties in quantum field theory.
\end{abstract}

\maketitle

\section{Introduction}

In quantum field theory (QFT), the interaction vertex between a single photon and a spin-$\frac{1}{2}$ fermion can be characterized in terms of four electromagnetic (EM) form factors.  In the limit of vanishing momentum transfer between the photon and fermion, the form factors encode the static EM properties of the particle, namely, its charge, magnetic dipole moment, electric dipole moment, and anapole moment.\cite{bk82,nieves} For beginning graduate students in QFT, two of these moments are certainly familiar, and likely, these students have seen how, in the low-energy limit, one can establish a connection between the QFT formalism and simple Coulombic potentials or magentic dipole interactions.  In this paper, we will focus our attention upon the anapole moment - probably the least familiar static property.  Classically, the anapole moment arises as a contact quadrupole moment in a multipole expansion of the magnetic field.\cite{gray_karl_novikov}   The paradigmatic example of a classical EM source with non-zero anapole moment is a toroidal current distribution; this paradigm carries over to the quantum realm.  Using the formalism of QFT, we will effect a scattering calculation that, in the low-energy limit, will make clear that the EM analogue of the quantum anapole is a point-like toroidal solenoid. 

The possibility of the anapole interaction was first noted by Zel'dovich \cite{zeldovich} after it was suggested that parity symmetry might be violated in the weak interactions.\cite{lee_yang}  He posited that the anapole moment $\mathbf{a}$, collinear with the particle's spin, interacted only with currents, $\mathbf{J}$, and not electric or magnetic fields.  The low-energy interaction Hamiltonian was thus $H_\text{int} = - \mathbf{a} \cdot \mathbf{J}$, which prompted the analogy between the anapole and a classical toroidal solenoid.  
 Experimental confirmation of parity violation \cite{wu} meant that the anapole moment was not a mere theoretical exercise, and decades later, the nuclear anapole moment of  cesium-133 was definitively extracted from a nearly dominant background of other interactions.\cite{Wood1759}

In contrast with the rich structure of a cesium nucleus, the sole static EM property of a spin-$\frac{1}{2}$ Majorana fermion is the anapole moment.\cite{kobzarev,bk82,nieves} Majorana fermions are self-conjugate fields; that is, there is no distinction between particle and antiparticle.  This fact severely constrains their electromagnetic interactions; they must be electrically neutral with vanishing electric and magnetic dipole moments.  
In terms of the Standard Model (SM) of particle physics, neutrinos are the only possible  Majorana fermions.  Experiments are underway to search for neutrinoless double-beta decay which, if observed, would confirm the neutrino's self-conjugate nature.\cite{exo200}  Moving beyond the SM, theories are rife with Majorana fermions.  For example, in supersymmetry, the fermionic partners of the neutral SM gauge and Higgs bosons are manifested as Majorana fermions known as neutralinos.  If these particles are the lightest supersymmetric particle, then they are a natural candidate for the dark matter (DM) of the universe.\cite{susy_dm}  
Outside of supersymmetry, there are other models of DM particles whose primary interaction is through the anapole moment,\cite{Fitzpatrick:2010br,anapole_dm1,anapole_dm2,DelNobile:2014eta} so their EM properties are germane to understanding, for instance, the relic density of DM.

In what follows, we will discuss, within the framework of QFT, the interaction vertex between a general spin-$\frac{1}{2}$ fermion and a single photon.  Then, narrowing our consideration to Majorana fermions, we will compute the scattering amplitude for a (spinless) charged scalar particle on a spin-$\frac{1}{2}$ Majorana fermion, which proceeds via the exchange of a single (virtual) photon.  In the limit of small momentum transfer, the interaction probes the anapole moment of the fermion. At such low energies, this amplitude should be equivalent to the quantum mechanical scattering amplitude for scalar particles passing near an electromagnetic target, described, in this case, by a particular magnetic vector potential.   We will find that the vector potential is consistent with a point-like toroidal solenoid.  Outside of this point-like solenoid, the magnetic field vanishes, but the vector potential does not.  As such, our scattering calculation can be used to explore the Aharonov-Bohm effect, a quantum mechanical phenomenon in which EM potentials impact the trajectory of the particle in a region with vanishing EM fields.  We find that our QFT calculation is consistent with the QM scattering for a spinless charged particle on an infinitesimally thin toroidal solenoid.

\section{Photon-fermion vertex}

\begin{figure}
\includegraphics[width=4.3cm]{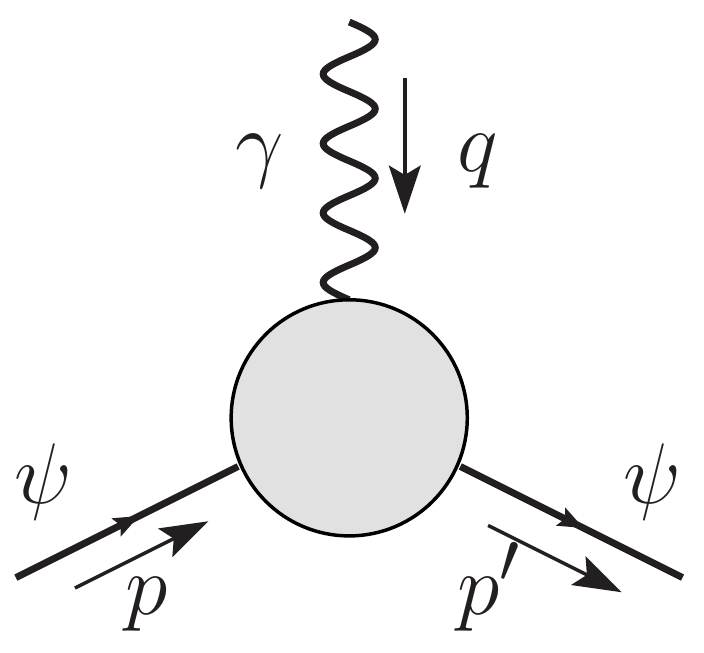}
\caption{ Interaction vertex between an on-shell fermion and an off-shell photon of momentum $q$. \label{fig1}}
\end{figure}

In QFT, the single-photon EM interactions of a spin-$\frac{1}{2}$ fermion are governed by a Lagrangian term consisting of a contraction between the fermion current $J^\mu_\text{EM} =\bar{\Psi} \Gamma^\mu \Psi$ and the photon vector field $A_\mu$.  By requiring the current to be Lorentz covariant and the overall interaction to be gauge invariant, one can decompose the operator $\Gamma^\mu$ into four distinct terms; a pedagogical treatment of this procedure can be found in Ref.~\onlinecite{em_formfactors}.  In the end, the matrix element for the electromagnetic current  characterizing the fermion transition from momentum and spin states $|\mathbf{p},s\rangle$ to $|\mathbf{p}',s'\rangle$ is given by 
\begin{equation}
\langle \mathbf{p}',s' | J_\text{EM}^\mu | \mathbf{p},s\rangle =  \bar{u}^{s'}(p')\left[f_1(q^2) \gamma^\mu + \frac{i}{2m} f_2(q^2)  \sigma^{\mu \nu} q_\nu + f_a(q^2) \left( q^2 \gamma^\mu -  q^\mu \qslash \right) \gamma^5 + f_e(q^2) \sigma^{\mu\nu} q_\nu \gamma^5  \right] u^s(p),  \label{vertex}
\end{equation}
where $q=p'-p$ is the four-momentum transferred to the fermion by the photon.\cite{bk82,nieves}  The only non-trivial scalar in this process is $q^2$, and thus the electromagnetic form factors are functions of this scalar.  In a Feynman diagram, this vertex will be represented with a shaded circle as in Fig.~\ref{fig1}.

In the limit of vanishing momentum transfer, i.e., $q^2 \to 0$, the four form factors encode the static EM properties of the fermion.   The electric charge of the particle is given by $f_1(0)$, and the anomalous magnetic dipole moment is related to $f_2(0)$.  For the remaining terms, $f_a(0)$ is the particle's anapole moment, and $f_e(0)$ is its electric dipole moment.  If a particle is to be its own antiparticle, then the particle must clearly be neutral, but it is not so obvious as to why a Majorana fermion's magnetic and electric dipole moments must vanish.  A rigorous field theoretic proof of this fact can be found in Refs.~\onlinecite{bk82,nieves}. 

These two papers also contain a more intuitive argument based upon the non-relativistic limit of the Hamiltonian which governs the interaction between the fermion and EM field.\cite{bk82,nieves}  The interaction Hamiltonian can be constructed by contracting the fermion EM current, Eq.~(\ref{vertex}), with the EM four-vector potential, $A^\mu$.  The factors of the photon's four-momentum, $q^\mu$, that appear in the EM fermion current are mapped to derivatives, $\partial^\mu$, of the four-vector potential in position space.  Heuristically, we  see that the  $f_2$ and $f_e$ terms in the Hamiltonian will involve the EM field strength tensor $F^{\mu \nu}\equiv \partial^\mu A^\nu - \partial^\nu A^\mu$, whereas the $f_a$ term will involve a derivative of the field tensor which can be related to a four-current via Maxwell's equations, $\partial_\mu F^{\mu\nu} = J^\nu$.  In the non-relativistic limit, the Hamiltonian takes the form $H_\text{int} = - \boldsymbol{\mu}\cdot \mathbf{B} - \mathbf{d} \cdot \mathbf{E} - \mathbf{a} \cdot \mathbf{J}$, at leading order.  

In a local QFT, the Hamiltonian must be invariant under a $\mathcal{CPT}$ transformation; that is, it must be unchanged when, jointly,  particles are changed to antiparticles, spatial axes are flipped, and time is reversed.
Each of the electromagnetic moments (magnetic dipole, $\boldsymbol{\mu}$; electric dipole, $\mathbf{d}$; and anapole, $\mathbf{a}$) is given by the product of a scalar and the fermion's spin.  As such, for a Majorana fermion, these moments pick up a minus sign under $\mathcal{CPT}$.  The electric and magnetic fields are even under $\mathcal{CPT}$ so that, overall, these magnetic and electric dipole terms in the Hamiltonian pick up an overall minus sign under $\mathcal{CPT}$.  As a result, the dipole moments must vanish.  On the other hand, the current density $\mathbf{J}$ is odd under $\mathcal{CPT}$ so that the anapole term in the Hamiltonian is indeed unchanged under the transformation.  In what follows, we will consider only Majorana fermions, setting $f_1, f_2, f_e \equiv 0$.

\section{Low-energy scattering}

As we mentioned above, real photons do not couple to the anapole moment. This can easily be ascertained from the operator for the anapole moment, $f_a(q^2) \left( q^2 \gamma^\mu -  q^\mu \qslash \right) \gamma^5 $.
Because a real photon is massless, the square of its momentum vanishes, $q^2=0$, and because it is transverse, the contraction of its polarization and momentum vectors vanishes, $\epsilon \cdot q=0$.  With these facts, it is clear that the anapole vertex operator will vanish when coupled to a real photon. As such, the only EM probe of an anapole is the EM current of some other particle that scatters off the fermion.

In QFT, the simplest current with which to probe the anapole is that produced by a scalar (spinless) particle of charge $e$ and mass $m_\phi$.  Using a scalar particle  lets us focus on the physics in the calculations without having spin complications in the current.
Given a scalar field $\phi$, the flow of charge can be described by the vector $J^\mu = i\frac{e}{2m_\phi} [\phi^* \partial^\mu \phi -(\partial^\mu \phi^*) \phi ]$.  Supposing the scalar particle is a plane wave with definite momentum $k^\mu$,  then $\phi \sim e^{-i k \cdot x}$ yields a current $J^\mu \sim \frac{e}{m_\phi} k^\mu $.

\begin{figure}
\includegraphics[width=4.3cm]{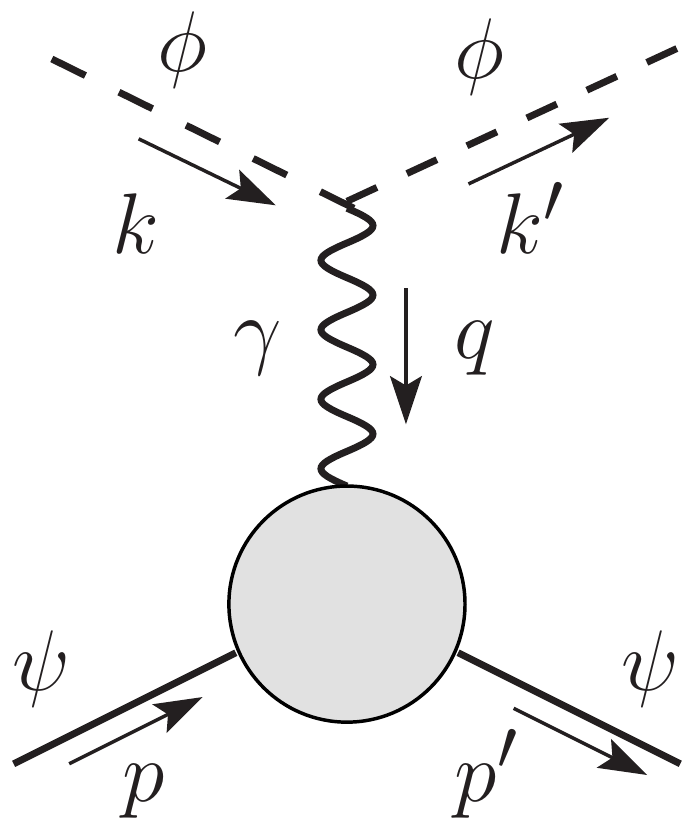}
\caption{ An incoming scalar, $\phi$, with momentum $k$ scatters off a Majorana fermion, $\psi$, with momentum $p$ via an anapole interaction.  The outgoing momenta for the scalar and fermion are $k'$ and $p'$, respectively.   \label{fig_feyndiag}}
\end{figure}

We compute the scattering amplitude for this scalar particle incident upon a Majorana fermion of mass $M_\psi$ and initial momentum $p$. The outgoing momenta of the scalar and fermion are $k'$ and $p'$, respectively.  Scattering is achieved by the exchange of a virtual photon of momentum $q\equiv p'-p$.  The fermion only interacts through its anapole moment.  The Feynman diagram for this scattering process is depicted in Fig.~\ref{fig_feyndiag}.
Following the conventions in Ref.~\onlinecite{peskin},   the amplitude for this  process is
\begin{equation}
\mathcal{M} = e f_a (k'_\mu + k_\mu) \bar{u}^{s'}(p') \left[\frac{(q^2 \gamma^\mu -q^\mu \qslash) \gamma^5}{q^2}  \right] u^s(p). \label{amp1}
\end{equation}
A remark is in order about the practical aspects of computing  scattering amplitudes for processes  involving Majorana fermions.  Majorana fermions are special solutions of the Dirac equation.  Using a particular (Majorana) representation of the Dirac matrices, Majorana fermions are manifestly real  with half the number of independent parameters as a Dirac fermion.  (This notion of ``realness" for a Majorana fermion can be generalized to an arbitrary representation of the Dirac matrices.\cite{pal})  The Feynman rules appearing in most QFT texts are appropriate for Dirac fermions, but with some modifications, one can adapt these same rules for Majorana fermions as in, for example,  Refs.~\onlinecite{denner1,denner2}.

Because Majorana fermions are solutions to the Dirac equation, the spinors satisfy $\pslash u^s(p) = M_\psi u^s(p)$ and $ \bar{u}^{s'}(p') \pslash'= M_\psi \bar{u}^{s'}(p')$. Armed with these identities, we can simplify the amplitude, Eq.~(\ref{amp1}), considerably:
\begin{equation}
\mathcal{M} = 2e f_a \bar{u}^{s'}(p') \left[ \kslash - \frac{2 ( k\cdot q) }{q^2}M_\psi  \right]  \gamma^5u^s(p). \label{amp2}
\end{equation}
This expression is entirely general; however, we wish to focus upon the limit in which the four-momentum transfer $q$ is small, so in what follows we will expand our expression in powers of this momentum transfer, keeping only the leading order terms.  Additionally, we may as well work within a specific frame - the rest frame of the fermion target.  

If we assume the scatterer to be at rest, then its four-momentum is  $p = (M_\psi, \mathbf{0})$. The scattered fermion's 3-momentum is that transferred by the photon, $\mathbf{q}$, so that $p' = (E', \mathbf{q})$ with $E' = \sqrt{M_\psi^2 +|\mathbf{q}|^2}$. Denoting the photon's energy as $q^0$, the energy of the scattered fermion can be written as $E' = M_\psi + q^0 = \sqrt{M_\psi^2 + | \mathbf{q}|^2}$.  Working with these two expressions for $E'$, we find that, in the limit of small momentum transfer, $q^0 \approx \frac{|\mathbf{q}|^2}{2 M_\psi}$;  this is just the non-relativistic kinetic energy of the scattered fermion.  With these approximations, we can estimate the explicit kinematical factor that appears in the amplitude, Eq.~(\ref{amp2}), as
\begin{equation}
\frac{k\cdot q}{q^2} \approx \frac{\mathbf{k}\cdot \mathbf{q}}{|\mathbf{q}|^2}-\frac{k^0}{2M_\psi}  + \mathcal{O}(|\mathbf{q}|).
\end{equation}

Now we evaluate the Dirac bilinears.  We note that a Dirac bispinor can be written in terms of Pauli matrices and two-component spinors $\xi^s$ as follows
\begin{equation}
u^s(p) = \frac{1}{\sqrt{2(p^0+M_\psi)}} \left(\begin{array}{c} [(p^0+M_\psi)\mathbbm{1} -\mathbf{p}\cdot \boldsymbol{\sigma} ] \xi^s \\  \left[ (p^0+M_\psi)\mathbbm{1} +\mathbf{p}\cdot \boldsymbol{\sigma} \right]  \xi^s \end{array}\right).
\end{equation}
With this expression for the bispinor and our small $q^2$ assumption, we approximate the bilinears 
\al{
\bar{u}^{s'}(p') \gamma^5 u(p) &\approx - {\xi^{s'}}^\dagger [\mathbf{q}\cdot \boldsymbol{\sigma}] \xi^s + \mathcal{O}(| \mathbf{q}|^3), \\
\bar{u}^{s'}(p') \gamma^0 \gamma^5 u(p) &\approx  {\xi^{s'}}^\dagger [\mathbf{q}\cdot \boldsymbol{\sigma}] \xi^s + \mathcal{O}(| \mathbf{q}|^3), \\
\bar{u}^{s'}(p') \gamma^j \gamma^5 u(p) &\approx  2M_\psi {\xi^{s'}}^\dagger\sigma^j \xi^s + \mathcal{O}(| \mathbf{q}|^2).}
With these approximations, we can then determine the leading order contributions to the scattering amplitude for small momentum transfer to be
\begin{equation}
\mathcal{M} = 8e M_\psi f_a  \left[  -\mathbf{k}\cdot \mathbf{S} + \frac{(\mathbf{k}\cdot \mathbf{q})( \mathbf{q}\cdot \mathbf{S})}{|\mathbf{q}|^2} \right] +\mathcal{O}(|\mathbf{q}|^2), \label{amp_lowq}
\end{equation}
where, for shorthand, we define the spin matrix element $\mathbf{S} = \frac{1}{2} {\xi^{s'}}^\dagger \boldsymbol{\sigma} \xi^s$.

Modulo some dimensional normalization factors, the scattering amplitude is proportional to the matrix element for the interaction Hamiltonian mediating the transition from initial to final states, $H_\text{f\,i} \sim \mathcal{M}$.  At low energies, the field theoretic amplitude should correspond to the quantum mechanical scattering amplitude.  Examining the approximate amplitude, Eq.~(\ref{amp_lowq}), the interaction Hamiltonian has the structure of the scalar's current interacting with the magnetic vector potential associated with the Majorana fermion, $H_\text{f\,i} \sim - e\mathbf{k} \cdot \mathbf{A}_a$.  Thus, the vector potential associated with the anapole moment of the Majorana fermion is
\begin{equation}
\mathbf{A}_a (\mathbf{q}) \sim f_a  \left[   \mathbf{S} - \frac{( \mathbf{q}\cdot \mathbf{S})}{|\mathbf{q}|^2}\mathbf{q} \right], \label{A_q}
\end{equation}
again, modulo some dimensional factors.
We note that both terms in Eq.~(\ref{A_q}) are both zeroth order in $|\mathbf{q}|$ in the Taylor expansion about the momentum transfer, despite the $\mathbf{q}$-dependence of the second term. 

The above relation for the vector potential, Eq.~(\ref{A_q}), is expressed in terms of its momentum representation.  By taking its inverse Fourier transform, we can determine the expression for the vector field in position space:
\al{
\mathbf{A}_a (\mathbf{r}) = & \frac{1}{(2\pi)^3} \int \mathbf{A}_a (\mathbf{q}) e^{i \mathbf{q}\cdot \mathbf{r}} \mathrm{d}^3 q \\
\sim& f_a  \left\{   \delta^{(3)} (\mathbf{r}) \mathbf{S} + \frac{1}{4\pi} \frac{1}{r^3}[ 3 (\mathbf{S}\cdot \hat{\mathbf{r}}) \hat{\mathbf{r}}- \mathbf{S}  ] \right\} \label{A_r}.
}
The first term in Eq.~(\ref{A_r}) arises from the $q$-independent term in Eq.~(\ref{A_q}).  This is the vector potential associated with a classical point-like anapole moment,\cite{gray_karl_novikov} with its moment aligning with the particle's spin vector.  The impact of this vector potential is confined to the origin.  The second term arises from the $\mathbf{q}$-dependent term in Eq.~(\ref{A_q}), and it extends through all of space.  We note that the structure of this second term is consistent with the large $r$ behavior of the vector potential (in the Coulomb gauge) outside of a  classical toroidal solenoid, again with the solenoid's symmetry axis lying along the direction of the particle's spin.\cite{afanasiev_dubovik}

To be concrete, we now suppose that the spin of the Majorana fermion is aligned with the $z$-axis and that no spin flip occurs during scattering, $s'=s$. Then, we have $\mathbf{S} = S \hat{\mathbf{z}}$ with $S=\frac{1}{2}$.  We can  take the curl of the vector potential to determine the associated magnetic field:
\begin{equation}
\mathbf{B} = \boldsymbol{\nabla} \times \mathbf{A} \sim -  f_a S \left( \frac{\partial }{\partial r} \delta^{(3)} (\mathbf{r})\right) \sin \theta \hat {\boldsymbol{\phi}}. \label{B_ana}
\end{equation}
The second term in Eq.~(\ref{A_r}) has no curl (nor divergence) and thus yields no contribution to the magnetic field.  This is consistent with the notion that the anapole moment of a Majorana fermion can be thought of as a point-like toroidal magnetic field.  The magnetic field in Eq.~(\ref{B_ana}) is only nonzero at the origin, but if smeared out, it would be directed azimuthally, also consistent with a torus whose axis lies on the $z$-axis.  The magnetic field outside the torus would vanish, as does the field for the anapole moment.  

Up to this point, we have assumed that the momentum transfer to the Majorana fermion was small relative to its mass, $M_\psi$.  Let us make some further simplifying assumptions. Suppose  that the charged scalar particles are also non-relativistic, i.e., $|\mathbf{k}| \ll m_\phi$. Additionally, we will take the target fermion mass to be much greater than the scalar particle's mass, $M_\psi \gg m_\phi$.  In this limit, the target recoil is negligible, and the scalar particle's kinetic energy is essentially unchanged $|\mathbf{k}| \approx |\mathbf{k}'|$.  Finally, we will align the incoming scalar current with the fermion's spin along the $z$-axis, $\mathbf{k}= |\mathbf{k}|\hat{\mathbf{z}}$.

With these assumptions, the scattering amplitude is approximated by
\begin{equation}
\mathcal{M} \approx -4 e f_a M_\psi S |\mathbf{k}| (1+ \cos \theta), \label{nr_m}
\end{equation}
where the scattering angle $\theta$ is the usual polar angle. 
From the amplitude we compute the differential cross section 
\begin{equation}
\frac{\mathrm{d} \sigma}{\mathrm{d} \Omega} = \frac{\alpha}{\pi} f_a^2 S^2 |\mathbf{k}|^2 (1+\cos \theta)^2, \label{nr_xsec}
\end{equation}
with $\alpha$ the fine structure constant.

\section{Application to the Aharonov-Bohm effect}

At low energies, the EM analogue of a Majorana fermion is a point-like toroidal solenoid with a magnetic field confined to the origin but a nonzero vector potential throughout all space.  Given this, we can use the previous calculation to explore the impact of the curl-free vector potential upon the scattering amplitude - a variant of the Aharonov-Bohm (AB) effect as applied to toroidal solenoids.  In their seminal paper, Aharonov and Bohm showed, theoretically, that in quantum mechanics the electron state can be measurably impacted by a magnetic vector potential even if the magnetic field vanishes along the electron's trajectory.\cite{AB}  
One of their proposed experimental tests was to have an electron beam pass on both sides of a solenoid with a confined magnetic field. The phase of an electron wave function changes passing the solenoid, with a net phase change between that part of the wave that passed on one side compared to the part of the wave that passed on the other side given by $\pm e\oint \mathbf{A} \cdot  \mathrm{d}\mathbf{r}$ for a line-integration path surrounding the solenoid.
This effect was confirmed experimentally,\cite{AB_exp} but some critics, e.g., Ref.~\onlinecite{roy}, claimed that fringing magnetic fields around the finite solenoid  were responsible for the observed phase shifts.  To counter this criticism, it was suggested that the effect be explored with a toroidal solenoid;\cite{AB_torus} with this geometry, the magnetic flux is completely contained within a region inaccessible to the electron.  Years later, the toroidal analogue of the original AB experiment yielded results consistent with the AB effect.\cite{AB_torus_exp}

Returning to the last section, we find that the non-relativistic differential cross section, Eq.~(\ref{nr_xsec}), strongly depends upon the vector potential for $r>0$ where the magnetic field is zero.  The angular dependence of the differential cross section  is derived wholly from the second term in the vector potential in Eq.~(\ref{A_r}).  If the only contribution to the scattering were due to the nonzero magnetic field, then the amplitude would be $\mathcal{M} =8 e f_a M_\psi S |\mathbf{k}|$ with, for example, forward scattering being equally as likely as backward scattering.  The physical consequences of the vector potential where the magnetic field vanishes is a pure quantum effect.

Because our calculation is entirely non-relativistic, we should be able to compare our QFT result with a non-relativistic quantum mechanical calculation of the scattering amplitude of a spinless charged particle incident upon an infinitesimal toroidal solenoid.  
In the literature, we find several such calculations.\cite{AB_torus, afanasiev1, afanasiev2} 
Focusing on Ref.~\onlinecite{AB_torus}, the authors compute this amplitude, assuming a thin classical toroidal solenoid of radius $R$ carrying a magnetic flux $\Phi$, in the first Born approximation. The authors also assume that the electron current and the solenoid's axis are both aligned with the $z$-axis, analogous to our above computation.
Converting their results to natural units,  they find 
\begin{equation}
f(\theta) = \frac{ 1}{4} e|\mathbf{k}| R^2 \Phi (1+\cos \theta),  \label{qm_scatt1}
\end{equation}
with a differential cross section given by $\diff \sigma/\diff \Omega = |f(\theta)|^2$.

This result is not expressed in terms of the anapole moment, so to aid comparison, we first compute the anapole moment for a classical ideal torus centered at the origin, with symmetry axis aligned with the $z$-axis.  We choose the current orientation so that the magnetic field is in the $\hat{\boldsymbol{\phi}}$ direction.  The simplest solenoid to consider is one with a rectangular cross-sectional area.  Let the height of the solenoid be $h$ and its interior and exterior radii be $R$ and $R+\epsilon$, respectively.  Supposing a current $I$ runs through $N$ turns of the solenoid, the magnetic field in the interior of the solenoid is $\mathbf{B} =  \mu_0 N I/(2\pi \rho) \hat{\boldsymbol{\phi}}$, where $\rho$ is the distance from the $z$-axis.  
Given this field, we compute the flux through the solenoid to be 
\begin{equation}
\Phi = \frac{\mu_0 N I}{2\pi} h  \ln \left[1 + \frac{\epsilon}{R} \right] \approx \frac{\mu_0 N I A}{2\pi R},
\end{equation}
where we assume a thin torus with $\epsilon \ll R$  and denote the torus's cross-sectional area as $A$.  
Adapting the definition of the anapole moment for a line current,\cite{gray_karl_novikov,note} we find
\al{
\mathbf{a} =& -\int I r^2 \mathrm{d} \boldsymbol{\ell}, \\
=&  NI [(R+\epsilon)^2-R^2] h \hat{\mathbf{z}},\\
\approx& 2 N I A R \hat{\mathbf{z}},
}
where $\mathrm{d}\boldsymbol{\ell}$ is a current-directed infinitesimal line element.  We note that contributions to the anapole moment from the radial current vanish.
Given this, we can rewrite the flux in terms of the magnitude of the anapole moment as $\Phi =  a/(4\pi R^2)$, where we move to natural units by setting $\mu_0 = 1$.  

In terms of the classical anapole moment, the quantum mechanical scattering amplitude in Eq.~(\ref{qm_scatt1}) can be written as
\begin{equation}
f(\theta) =  e |\mathbf{k}|  \frac{a}{16\pi }  (1+\cos \theta).
\end{equation}
Given our choice of conventions, the quantum mechanical amplitude is related to the QFT amplitude in Eq.~(\ref{nr_m}) via $f(\theta) = \mathcal{M}/(8\pi M_\psi)$.  Comparing the two, we see that both amplitudes will result in the same differential cross section assuming  we identify the Majorana fermion's anapole moment with the classical one via $a = 8 f_a S$.

\section{Conclusion}

Because a Majorana fermion is self-conjugate, its sole static electromagnetic property is the anapole moment.  In single-photon interactions, we have seen that these particles only couple to virtual (but not real) photons.  This corresponds to the notion that anapoles only experience interactions due to the currents of other particles, but not to local external electric or magnetic fields.  Through low-energy scattering, we see that the quantum anapole behaves like a point-like toroidal solenoid. Assuming the fermion's spin is aligned with the $z$-axis, the anapole's magnetic field is azimuthal in direction and confined to the origin. Because the anapole's magnetic vector potential is nonzero throughout space, one can use a Majorana fermion to explore the AB effect with toroidal geometry.

In our scattering calculations, we have exclusively explored the low-energy limit, assuming small momentum transfer.  But, because we work in a field theoretic framework, one could easily explore the behavior of the anapole moment in the relativistic regime and generalize to currents from particles carrying spin.   The only word of caution in extending the discussion to higher energies is to note that the anapole moment represents an effective interaction; that is, the coupling between a Majorana fermion and photon is mediated by charged particles in a more fundamental theory.  
If one (or both) of these charged particles is more massive than the Majorana fermion, then its stability is assured.
If the dominant-mass particle is {\it much} heavier than the Majorana fermion, then the details of these charged-particle interactions are irrelevant at low energies.
But, as the momentum transfer in a scattering process increases, further underlying physics could become appreciable.  As an example, suppose we were to compute the real Compton amplitude for a Majorana fermion, which involves the interaction between two real photons (viz., the incident and subsequently scattered photons)  and the fermion.
We would conclude that the amplitude vanishes because anapoles do not couple to real photons. 
However, generically, the coupling between Majorana fermions and {\em two} photons is nonzero.\cite{maj_2photon} So, at sufficient energies, Majorana fermions {\em can} scatter photons.

\section{Acknowledgment}
KMW was funded, in part, by a McCormick Scholar award from the University of Puget Sound.
DCL was funded, in part, by a Mellon Junior Sabbatical Fellowship and a McCormick Faculty Mentor award from the University of Puget Sound.


\begin{thebibliography}{99}

\bibitem{bk82} 
Boris Kayser, ``Majorana neutrinos and their electromagnetic properties,'' Phys. Rev. D {\bf 26}, 1662 (1982).

\bibitem{nieves}
Jose F. Nieves, ``Electromagnetic properties of Majorana neutrinos,'' Phys. Rev. D {\bf 26}, 3152 (1982).


\bibitem{gray_karl_novikov}
 C. G. Gray, G. Karl, and V. A. Novikov, ``Magnetic multipolar contact fields: The anapole and related moments,'' Am. J. Phys. {\bf 78}, 936--948 (2010).


\bibitem{zeldovich}
 Ya. B. Zel'dovich, ``Electromagnetic interaction with parity violation,'' Sov. Phys. JETP {\bf 6}, 1184 (1957), [Zh. Eksp. Teor. Fiz. {\bf 33}, 1531 (1957)].

\bibitem{lee_yang} 
T. D. Lee and C. N. Yang, ``Question of parity conservation in weak interactions,'' Phys. Rev. {\bf 104}, 254--258 (1956).

\bibitem{wu} 
C. S. Wu, E. Ambler, R. W. Hayward, D. D. Hoppes, and R. P. Hudson, ``Experimental test of parity conservation in beta decay,'' Phys. Rev. {\bf 105}, 1413--1415 (1957).

\bibitem{Wood1759} 
C. S. Wood, S. C. Bennett, D. Cho, B. P. Masterson, J. L. Roberts, C. E. Tanner, and C. E. Wieman, ``Measurement of parity nonconservation and an anapole moment in cesium,'' Science {\bf 275}, 1759--1763 (1997).

\bibitem{kobzarev}
 I.Yu Kobzarev and L. B. Okun, in {\em Problems of Theoretical Physics} (Publishing House Nauka, Moscow, 1972) pp. 219--224.

\bibitem{exo200}
 J. B. Albert et al. (EXO-200), ``Search for Majorana neutrinos with the first two years of EXO- 200 data,'' Nature {\bf 510}, 229--234 (2014).

\bibitem{susy_dm}
 Gerard Jungman, Marc Kamionkowski, and Kim Griest, ``Supersymmetric dark matter,'' Phys. Rept. {\bf 267}, 195--373 (1996).

\bibitem{Fitzpatrick:2010br}
A. Liam Fitzpatrick and Kathryn M. Zurek, ``Dark moments and the DAMA-CoGeNT puzzle,'' Phys. Rev. D {\bf 82}, 075004 (2010).

\bibitem{anapole_dm1}
 Chiu Man Ho and Robert J. Scherrer, ``Anapole dark matter,'' Phys. Lett. B {\bf 722}, 341--346 (2013).

\bibitem{anapole_dm2}
Yu Gao, Chiu Man Ho, and Robert J. Scherrer, ``Anapole dark matter at the LHC,'' Phys. Rev. D {\bf 89}, 045006 (2014).

\bibitem{DelNobile:2014eta}
Eugenio Del Nobile, Graciela B. Gelmini, Paolo Gondolo, and Ji-Haeng Huh, ``Direct detection of light anapole and magnetic dipole DM,'' JCAP {\bf 1406}, 002 (2014).

\bibitem{em_formfactors}
 Marek Nowakowski, E. A. Paschos, and J. M. Rodriguez, ``All electromagnetic form-factors,'' Eur. J. Phys. {\bf 26}, 545--560 (2005).

\bibitem{peskin} 
Michael E. Peskin and Daniel V. Schroeder, {\em An Introduction to Quantum Field Theory} (Addison-Wesley, Reading, USA, 1995).

\bibitem{pal} 
  Palash B.~Pal,
  ``Dirac, Majorana, and Weyl fermions,''
  Am.~J.~Phys.~{\bf 79}, 485--498 (2011).

\bibitem{denner1}
  A.~Denner, H.~Eck, O.~Hahn and J.~Kublbeck,
  ``Compact Feynman rules for Majorana fermions,''
  Phys.\ Lett.\ B {\bf 291}, 278--280 (1992).

\bibitem{denner2}
  A.~Denner, H.~Eck, O.~Hahn and J.~Kublbeck,
  ``Feynman rules for fermion number violating interactions,''
  Nucl.\ Phys.\ B {\bf 387}, 467--481  (1992).


 
 \bibitem{afanasiev_dubovik}
   G.~N.~Afanasiev  and V.~M.~Dubovik, ``Some remarkable charge-current configurations,'' Phys. Part. Nucl. {\bf 29}, 366--391 (1998).


\bibitem{AB}
 Y. Aharonov and D. Bohm, ``Significance of electromagnetic potentials in the quantum theory,'' Phys. Rev. {\bf 115}, 485--491 (1959).

\bibitem{AB_exp}
 R. G. Chambers, ``Shift of an electron interference pattern by enclosed magnetic flux,'' Phys. Rev. Lett. {\bf 5}, 3--5 (1960).

\bibitem{roy}
 S. M. Roy, ``Condition for nonexistence of Aharonov-Bohm effect,'' Phys. Rev. Lett. {\bf 44}, 111--114 (1980).

\bibitem{AB_torus}
 V.~L.~Lyuboshitz and Y.~A.~{Smorodinski{\v i}}, ``The Aharonov-Bohm effect in a toroidal solenoid,'' Sov. Phys. JETP {\bf 48}, 19 (1978).

\bibitem{AB_torus_exp}
 Nobuyuki Osakabe, Tsuyoshi Matsuda, Takeshi Kawasaki, Junji Endo, Akira Tonomura, Shinichiro Yano, and Hiroji Yamada, ``Experimental confirmation of Aharonov-Bohm effect using a toroidal magnetic field confined by a superconductor,'' Phys. Rev. A {\bf 34}, 815--822 (1986).

\bibitem{afanasiev1} G.~N.~Afanasiev, ``The scattering of charged particles on the toroidal solenoid,'' J. Phys. A: Math. Gen. {\bf 21}, 2095--2110 (1988).

\bibitem{afanasiev2} G.~N.~Afanasiev, ``Theoretical description of Tonomura-like experiments (electron scattering on a toroidal solenoid),'' Phys. Lett. A {\bf 142}, 222--226 (1989).

\bibitem{note}
In our definition of the anapole moment, we introduce a minus sign relative to Ref.~\onlinecite{gray_karl_novikov} in order to ensure that a toroidal magnetic field in the $\hat{\boldsymbol{\phi}}$ direction yields an anapole moment in the $\hat{\mathbf{z}}$ direction. This choice is consistent with quantum treatments of the anapole moment, cf.~Ref.~\onlinecite{dubovik}.

\bibitem{dubovik}
V.~M.~Dubovik and A.~A.~Cheshkov, 
``Multipole expansion in classical and quantum field theory and radiation,'' Sov.~J.~Part.~Nucl.~{\bf 5}, 318--337 (1975).

\bibitem{maj_2photon} David C. Latimer, ``Two-photon interactions with Majorana fermions,'' Phys. Rev. D {\bf 94}, 093010 (2016).

\end{thebibliography}
\end{document}